# In-depth characterization and analysis of simple shear flows over regularly arranged micro pillars, II. Effect of pillar arrangement

Yanxing Wang[1,*], Hui Wan[2], Tie Wei[3], and Fangjun Shu[1],

[1]Department of Mechanical and Aerospace Engineering, New Mexico State University, Las Cruces, NM 88011, USA

[2]Department of Mechanical and Aerospace Engineering, University of Colorado Colorado Springs, Colorado Springs, CO 80918, USA

[3]Department of Mechanical Engineering, New Mexico Institute of Mining and Technology, Socorro, NM 87801, USA

Through high-fidelity numerical simulation, the effect of the arrangement of micropillars on the flow characteristics and momentum transport has been extensively investigated. The surface friction due to the complex flow characteristics and momentum transport mechanism has also been studied in depth. The micropillars are arranged in a quadrilateral, and different arrangements are acquired by changing the streamwise and spanwise distances between pillar rows. The results show that the streamwise and spanwise pillar distances have their own different influences. When the streamwise pillar distance is small, the micro eddies in the gaps between the streamwise neighboring pillars are significantly suppressed. The increase in the spanwise pillar distance enhances the momentum transport from the flow above pillar array to the flow in the spaces among micro pillars. When the spanwise pillar distance is small, the micro eddies in the gaps between the streamwise neighboring pillars connect with each other and form a tubular eddy between each pair of spanwise pillar rows. The tubular eddies significantly reduce the momentum transport from the upper flow to the lower flow. The increase in the streamwise pillar distance increase the momentum flux slightly. The surface friction can be decomposed into three components which are associated with two factors, the dilution effect of the number density of micro pillars and the multi-faceted effects of micro eddies. These two factors are determined by the streamwise and spanwise pillar distances. The dependence of the total friction and its components on the pillar distances has been thoroughly examined.

## I. INTRODUCTION

There are a large number of hierarchical structured surfaces in natural organisms, which have demonstrated many unique functions, such as self-cleaning, anti-adhesion, and low friction [1,2]. Inspired by the natural surfaces, various functional surfaces with embedded surface structures have been fabricated and broadly used as a means of flow control and thermal management. The applications include microscale heat exchangers [3,4], in vitro microfluidic platform for cellular mechanotransduction analysis [5,6], and hierarchical structured antifouling surfaces [7-9]. The flow characteristics cover a wide range of regimes, from creeping flows at very low Reynolds numbers to turbulent flows at high Reynolds numbers [10-12]. In the fundamental research and practical applications of fluid dynamics over micro-structured surfaces, the characterization of the fluid flow and the momentum transport process is one of the important topics. In particular, the mechanical force caused by the complex fluid dynamics has been widely concerned. For example, extracellular forces, such as flow shear stress, can mediate the endogenous actomyosin-based intracellular cytoskeleton contractile force, thereby regulating cell behavior [13]. In order to investigate the flow-mediated cellular mechanotransduction process, various microfluidic platforms with embedded surface structures have been established as in vitro models to control the cellular culture environment [5,6]. In microscale heat exchange systems, surface structures are widely used to improve heat transfer efficiency. The use of surface structures, however, also increase the pressure drop, i.e., increase the pumping energy loss. The trade-off between heat transfer rates and pressure drop is one of the core issues in the development of micro-structured heat exchangers [3,4]. Although a considerable amount of effort has been devoted to the fundamental research of flow dynamics over micro-structured surfaces, and some experimental measurements and numerical simulations have been carried out [14-16], the current understanding is still far from enough to reveal the complex mechanisms of momentum transport, and the precise control of flow evolution and surface friction is not really achieved.

---

*yxwang@nmsu.edu

In Ref. [17], we numerically studied the general characteristics of a simple shear flow over the regularly arranged microscale pillars and the effect of fluid inertia on flow evolution and surface friction. Above the pillar array, the streamwise velocity decreases linearly from the moving top plane to the plane at pillar height. Below the pillar tips, the velocity is significantly reduced and a microscale eddy is generated in each gap between the streamwise neighboring pillars. At smaller Reynolds number, the fluid inertia is weak and the flow patterns on the windward and leeward sides of micro pillars are symmetric about the pillar center. As the Reynolds number increases, the fluid inertia takes effect and breaks the symmetric flow pattern. The fluid inertia makes the overhead flow tilt downward, forming a spiral long-range advection between the fluid flow above pillar array and the flow in the spaces among micro pillars. The long-range advection modifies the characteristics of flow evolution and the momentum transport from the upper fluid to the fluid among micro pillars as well as to micro pillars. The surface drag is decomposed into three components, the reaction force of micro pillars due to flow shear stress on pillar surfaces, the reaction force of micro pillars due to flow pressure on pillar surfaces, and the reaction force of bottom plane due to flow shear on bottom surface. For larger Reynolds numbers, fluid inertia prevents the fluid from flowing along the curved surface of micro pillars and reduces the equivalent shear stress of the reaction force due to flow shear on pillar surface. At the same time, the fluid inertia makes the overhead flow impact the windward side of micro pillars more strongly and therefore increases the equivalent shear stress of the reaction force due to flow pressure on pillar surfaces.

The detailed flow characteristics and momentum transfer not only depend on the Reynolds number, but also on the aspect ratio and arrangement of micro pillars. The variation of pillar aspect ratio and arrangement will also change the shear stress and pressure on the surfaces of micro pillars and bottom plane, thereby changing the total surface friction. So far, the research on the effect of the geometry and arrangement of surface structures is rare. This causes the lack of key factors to achieve the precise flow control.

As an extension of the work in Ref. [17], this paper systematically investigates the effect of arrangement of micro pillars on the characteristics of a simple shear flow, especially the influence on the components of total surface friction. The objective of this paper is to establish a complete picture of flow characteristics and decomposed surface friction with respect to the regular arrangement of micro pillars. This work will provide important insights for the understanding of flow physics induced by surface structures and provide critical guidelines for realizing the precise control of flow over micro-structured surfaces. This paper is organized as follows. The physical model and numerical methods are presented in Section II. The results are analyzed in Section III, followed by the conclusions in Section IV.

## II. PHYSICAL MODEL AND NUMERICAL METHODS

The physical model is shown in Fig. 1, in which an incompressible flow is confined by two infinitely large parallel planes, with structured thin pillars with round tips lining the bottom plane. The coordinates are defined as streamwise ($x$), spanwise ($y$) and wall-normal ($z$), with corresponding velocity components as $u_x$, $u_y$ and $u_z$, respectively. The distance between the two planes is denoted by $H$. The bottom plane is fixed and the top plane moves at a constant speed denoted by $U_0$ to produce a flow shear. The length and diameter of the micro pillars are denoted by $l_p$ and $D_p$, respectively. The micro pillars are arranged in a quadrilateral. The streamwise and spanwise distances between neighboring pillar centerlines are denoted as $\delta_x$ and $\delta_y$, respectively. The shear rate of the bulk flow can be roughly estimated by

$$S = U_0/H. \tag{1}$$

The Reynolds number based on the bulk flow shear rate $S$ and the pillar length is

$$Re = Sl_p^2/v, \quad (2)$$

where $v$ is the kinematic viscosity of the fluid. In the analysis, the pillar length $l_p$ and equivalent velocity at pillar tips $U_t = Sl_p$ are used as the characteristic length and velocity to normalize the spatial coordinates and flow velocity,

$$\tilde{x} = x/l_p, \quad (3)$$

$$\tilde{\mathbf{u}} = \mathbf{u}/U_t, \quad (4)$$

In the normalized form, the pillar length is $\tilde{l}_p = 1$ and the equivalent velocity at pillar tips is $\tilde{U}_t = 1$. A complete description of the problem includes Reynolds number based on flow shear rate, $Re$, pillar aspect ratio, $\tilde{l}_p/\tilde{D}_p$, distance between two planes, $\tilde{H}$, and streamwise and spanwise distances between neighboring pillars, $\tilde{\delta}_x$ and $\tilde{\delta}_y$. In this study, $\tilde{l}_p/\tilde{D}_p$ is fixed at 5, and $\tilde{H}$ is fixed at 3. Therefore, the influencing parameters reduce to $Re$, $\tilde{\delta}_x$ and $\tilde{\delta}_y$. Two Reynolds numbers, $Re = 1$ and 33, which are typical in microfluidic systems with apparent advection effects, are considered in this study. The effects of the streamwise and spanwise pillar distances are the focus of this paper. The distance-to-diameter ratios, $\tilde{\delta}_x/\tilde{D}_p$ and $\tilde{\delta}_y/\tilde{D}_p$, range from 2 to 8, covering the main region with significant influence of pillar distance.

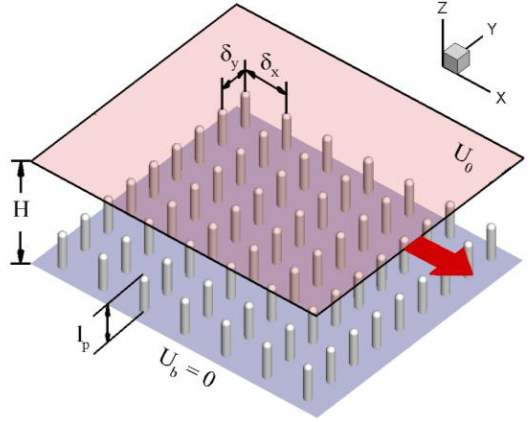

FIG. 1. Physical model of a simple shear flow with micro-pillar array on lower surface.

In this study, a multi-grid strategy within the lattice Boltzmann framework was used to simulate the incompressible fluid flow over the regularly arranged pillar array [17]. The D3Q15 model combined with a multi-relaxation-time (MRT) model was utilized with the consideration of minimizing the computational load in the low Reynolds number range [18,19]. In the treatment of non-slip conditions on solid boundaries, the scheme with 2nd order of accuracy was used at the pillar surfaces and the top and bottom planes [20]. To reduce the computational load from an exceptionally fine uniform grid throughout the domain as required by the LBM, a dual-lattice method was utilized [21], with a fine grid placed in the lower region surrounding the micro pillars. Within the fine grid, 25 grid points are used over the pillar diameter, which is enough to resolve the flow details considered in this study. The ratio of coarse grid size to fine grid size is $\delta x_c/\delta x_f = 5$, ensuring the smooth transition of flow quantities across the interface between the two grids. The height of the fine grid region is $\tilde{h}/\tilde{H} = 0.5$. The details are given in [17].

We selected several typical cases and ran the simulations over 10 structure elements in both streamwise and spanwise directions. The flow patterns demonstrate spatial periodicities same as the distances between micro pillars in both directions. Therefore, in the large-scale systematic study, only the flow within one cuboid domain including a single pillar was simulated, and periodic condition was utilized on the streamwise and spanwise surfaces. The analysis was carried out after the flow and scalar evolution entered a steady state in which all quantities remain constant over time.

The model has been extensively validated in our previous studies, and the details are described in Wang [17,22-24]. The grid independence study was given in [17]. The grid with 25 grid points over one pillar diameter was used in the simulation, for which the maximum deviation of the quantities from those using 35 points is less than 1%.

### III. RESULTS AND DISCUSSION

In this section, we will use several typical arrangements of micro pillars as examples to study the flow characteristics disturbed by pillar array. By qualitatively and quantitively comparing the detailed flow characteristics, the effect of pillar arrangement will be obtained. In particular, the effect of pillar arrangement on the decomposed surface friction will be analyzed in depth.

## *Basic Flow Characteristics*

In Ref. [17], it has been shown that the fluid flows at $Re_S = 1$ and 33 have similar patterns, except for the slight deviation caused by the fluid inertia at higher Reynolds numbers. Therefore, we will use the flows at $Re = 33$ as a representative to study the effects pillar arrangement on the basic characteristics of the flows. As a typical example, the flow with $\tilde{\delta}_x = \tilde{\delta}_y = 4\tilde{D}_p$ has been analyzed in [17]. To facilitate the discussion in this paper, a brief overview of that flow is given in Figs. 2 - 4. The flow patterns containing multiple pillars shown in the figure were generated by concatenating the patterns of one pillar. In the steady state, the streamlines coincide with the trajectories of fluid particles. As shown in Fig. 2(a), the finite thickness and height of micro pillars form a rectangular gap between each pair of streamwise neighboring pillars, in which the velocity magnitude is reduced. The flow above the gaps drives a clockwise recirculating eddy in each gap. Basically, the size of the eddies is determined by the geometry and arrangement of the pillars, as well as the Reynolds number. Induced by these eddies, the fluid surrounding the pillars takes an upward movement on the leeward side of the pillars and a downward movement on the windward side as it travels downstream. Figure 2(b) shows the patterns of streamwise velocity $\tilde{u}_x$ in the flow. Both the 3D iso-surfaces around the pillar array and 2D iso-contours on the streamwise plane show that the magnitude of streamwise velocity is significantly reduced in the flow layer below the pillar tips. Among the micro pillars the 3D iso-surface curves downward apparently, indicating the downward transport of the streamwise component of fluid momentum. Fig. 2(c) shows the patterns of wall-normal velocity ($\tilde{u}_z$) induced by pillar array. When the overhead flow climbs over the pillar tips, the upward and downward motions create a region with positive $\tilde{u}_z$ on the windward side and a region with negative $\tilde{u}_z$ on the leeward side of the hemispherical tips of micro pillars. Within the gaps between the streamwise pillars, $\tilde{u}_z$ is positive on the leeward side and is negative on the windward side of the pillars, due to the clock-wise recirculating eddy driven by the overhead flow (Fig. 2(a)). The magnitude of $\tilde{u}_z$ is smaller in the gaps, compared to that around the pillar tips, as shown in Fig. 2(c).

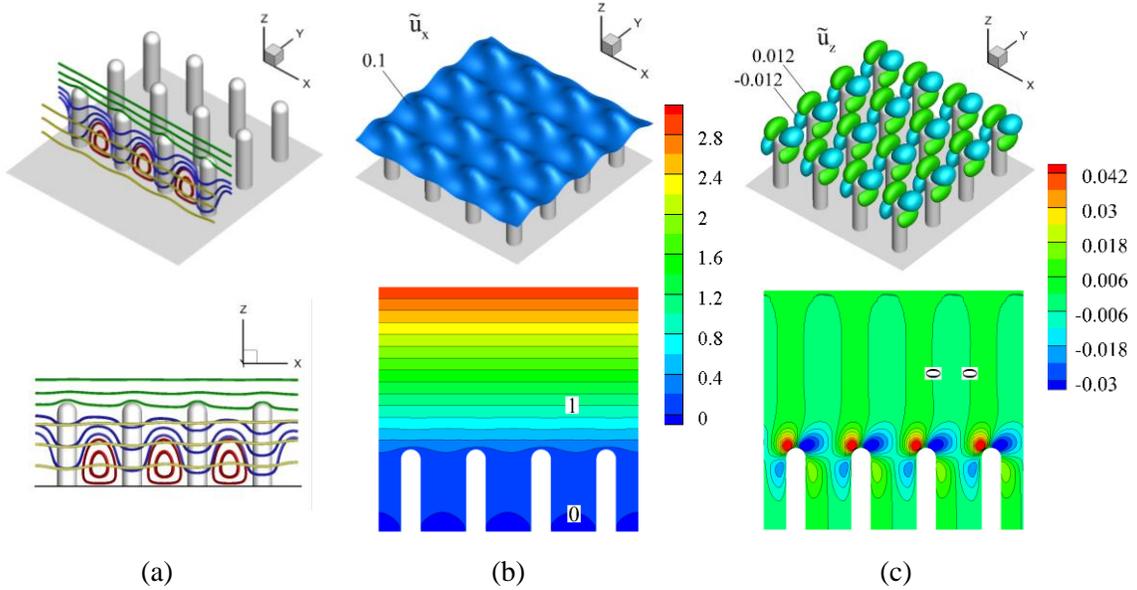

FIG. 2. Patterns of velocity characteristics for $\tilde{\delta}_x = \tilde{\delta}_y = 4\tilde{D}_p$. (a) 3D streamlines around pillars, (b) axial velocity $\tilde{u}_x$, and (c) vertical velocity $\tilde{u}_z$. The iso-contours are on the streamwise plane through pillar centers.

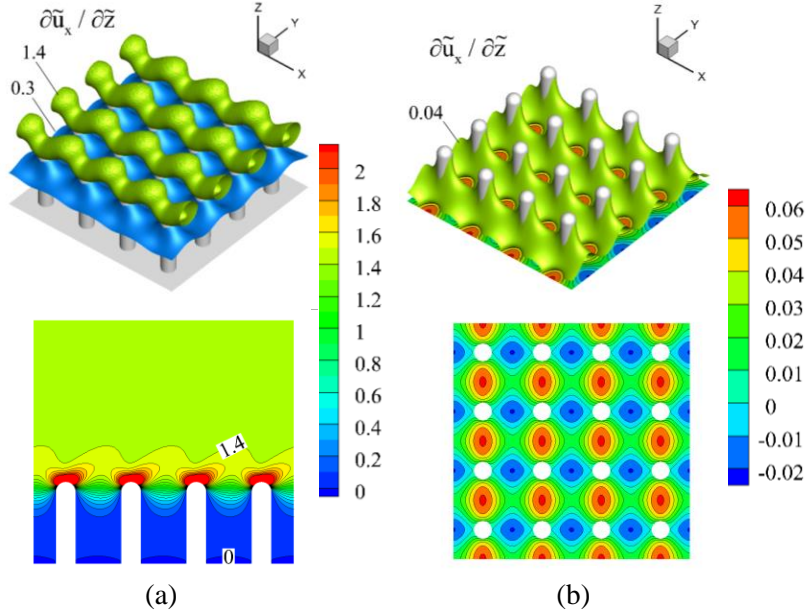

(a)             (b)

FIG. 3. Patterns of vertical gradient of axial velocity ($\partial \tilde{u}_x/\partial \tilde{z}$) for $\tilde{\delta}_x = \tilde{\delta}_y = 4\widetilde{D}_p$. (a) Entire regime, and (b) in the lower region.

For the simple shear flow considered in this study, the dominant component of the shear rate is the vertical gradient of streamwise velocity ($\partial \tilde{u}_x/\partial \tilde{z}$). Figure 3 shows the patterns of $\partial \tilde{u}_x/\partial \tilde{z}$ around the pillar array for $\tilde{\delta}_x = \tilde{\delta}_y = 4\widetilde{D}_p$. The micro pillars disturb the flow in the lower region and make the flow suddenly change direction around the pillar tips. This sudden change in flow direction causes an increase in $\partial \tilde{u}_x/\partial \tilde{z}$. The regions with increased $\partial \tilde{u}_x/\partial \tilde{z}$ around pillar tips extend in the streamwise direction and connect with each other, forming a curved tubular shape on each streamwise row of micro pillars, as shown by the 3D iso-surfaces in Fig. 3(a). In the spaces among micro pillars (Fig. 3(b)), the magnitude of $\partial \tilde{u}_x/\partial \tilde{z}$ reduces significantly, and $\partial \tilde{u}_x/\partial \tilde{z}$ shows a staggered pattern with positive and negative values on the bottom plane. Because of the clockwise recirculating eddies in the gaps between the streamwise pillars, $\partial \tilde{u}_x/\partial \tilde{z}$ is negative beneath the eddies.

Figure 4 shows the distributions of shear rate magnitude ($|\partial \tilde{u}_s/\partial n|$, where $\tilde{u}_s$ is the tangential velocity, and $n$ is the vector in normal direction) and pressure ($\tilde{p}$) at the pillar surface. As shown in the figure, larger $|\partial \tilde{u}_s/\partial n|$ and $\tilde{p}$ mainly appear at pillar tips due to the direct interaction with the overhead flow. The shear stress and pressure at the pillar surfaces provide the important components of surface friction on the structured surfaces.

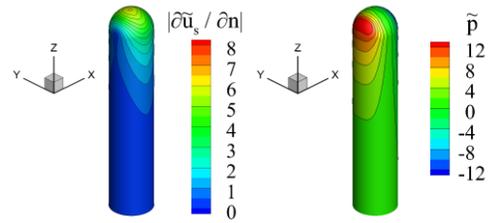

FIG. 4. Distribution of flow shear rate and pressure at the surface of a micro pillar for $\tilde{\delta}_x = \tilde{\delta}_y = 4\widetilde{D}_p$.

The basic flow characteristics for $\tilde{\delta}_x = \tilde{\delta}_y = 4\widetilde{D}_p$ discussed above will be the basis for the analysis of pillar arrangement effect below.

### *Effects of pillar arrangement on basic flow characteristics*

We first change the pillar distances in both the streamwise ($\tilde{\delta}_x$) and spanwise ($\tilde{\delta}_y$) directions at the same time and keep $\tilde{\delta}_x = \tilde{\delta}_y$. Figure 5 shows the patterns of 3D streamlines, streamwise velocity ($\tilde{u}_x$) and vertical velocity ($\tilde{u}_z$) for $\tilde{\delta}_x = \tilde{\delta}_y = 2\widetilde{D}_p, 4\widetilde{D}_p, 6\widetilde{D}_p$ and $8\widetilde{D}_p$. As shown in Fig. 5(a), the streamlines are characterized by a recirculating eddy within each gap between the streamwise neighboring pillars. The eddies first increase in size as $\tilde{\delta}_x$ and $\tilde{\delta}_y$ increase from $2\widetilde{D}_p$ to $4\widetilde{D}_p$, then decrease as $\tilde{\delta}_x$ and $\tilde{\delta}_y$ increase to $6\widetilde{D}_p$, and finally disappear as $\tilde{\delta}_x$ and $\tilde{\delta}_y$ increase to $8\widetilde{D}_p$. Basically, the eddy size is determined by the combination of streamwise ($\tilde{\delta}_x$) and spanwise ($\tilde{\delta}_y$) pillar distances, pillar diameter ($\widetilde{D}_p$), pillar length ($\tilde{l}_p$), and Reynolds number ($Re$). Generally, larger $\tilde{\delta}_x$, $\widetilde{D}_p$ and $\tilde{l}_p$ form larger confined areas between the streamwise neighboring pillars, thereby creating larger eddies. Yet larger $\tilde{\delta}_y$ induces larger fluid velocity in the area between the streamwise row of micro pillars, which will suppress the micro eddies. When $\tilde{\delta}_x$ and $\tilde{\delta}_y$ increase from $2\widetilde{D}_p$ to $4\widetilde{D}_p$, the increasing $\tilde{\delta}_x$ plays a dominant role, so the eddy size increases. When $\tilde{\delta}_x$ and $\tilde{\delta}_y$ increase from $4\widetilde{D}_p$ to $8\widetilde{D}_p$, the effect of increased $\tilde{\delta}_y$ becomes more important, which suppresses the eddies. Fig. 5(b) shows the 3D iso-surfaces and 2D iso-contours of the streamwise velocity ($\tilde{u}_x$). The 2D iso-contours are on the streamwise plane through pillar centers. The iso-surfaces curve downward among the pillars, and the degree of curving is larger for larger $\tilde{\delta}_x$ and $\tilde{\delta}_y$. This implies that as $\tilde{\delta}_x$ and $\tilde{\delta}_y$ increases, the momentum flux from the upper fluid to the fluid in the spaces among micro pillars increases, thereby the magnitude of $\tilde{u}_x$ in the spaces increases. The 2D contours confirm this conclusion. Fig. 5(c) shows patterns of vertical velocity ($\tilde{u}_z$). Around the pillar tips, a region with positive $\tilde{u}_z$ and a region with negative $\tilde{u}_z$ are formed on the windward and leeward sides of pillar tips, respectively, corresponding to the upward and downward movement as the overhead flow climbs over the hemispherical pillar tips. Within the gaps between streamwise neighboring pillars, the overhead flow attracts an upward flow on the leeward side of the pillar and pushes a downward flow on the windward side of the pillar. As a result, a region with negative $\tilde{u}_z$ and a region with positive $\tilde{u}_z$ are formed on the windward and leeward sides of pillar bodies, respectively. When $\tilde{\delta}_x$ and $\tilde{\delta}_y$ increases from $2\widetilde{D}_p$ to $8\widetilde{D}_p$, the increased momentum flux from the upper fluid to the fluid among micro pillars leads to stronger interaction between the fluid and the pillars, resulting greater velocity oscillation in vertical direction, around the pillar tips and in the spaces among micro pillars. Therefore, the magnitude of $\tilde{u}_z$ increases with the increase in $\tilde{\delta}_x$ and $\tilde{\delta}_y$. In addition, the influential area of the disturbed $\tilde{u}_z$, that is, the size of the iso-surfaces also increases, as shown in the figure. It should be noted that the increase in $\tilde{u}_z$ is not directly related to the size of micro eddies. When $\tilde{\delta}_x$ and $\tilde{\delta}_y$ increases from $4\widetilde{D}_p$ to $8\widetilde{D}_p$, the micro eddies gradually decrease in size and disappear as shown in Fig. 5(a), yet the magnitude of $\tilde{u}_z$ increases.

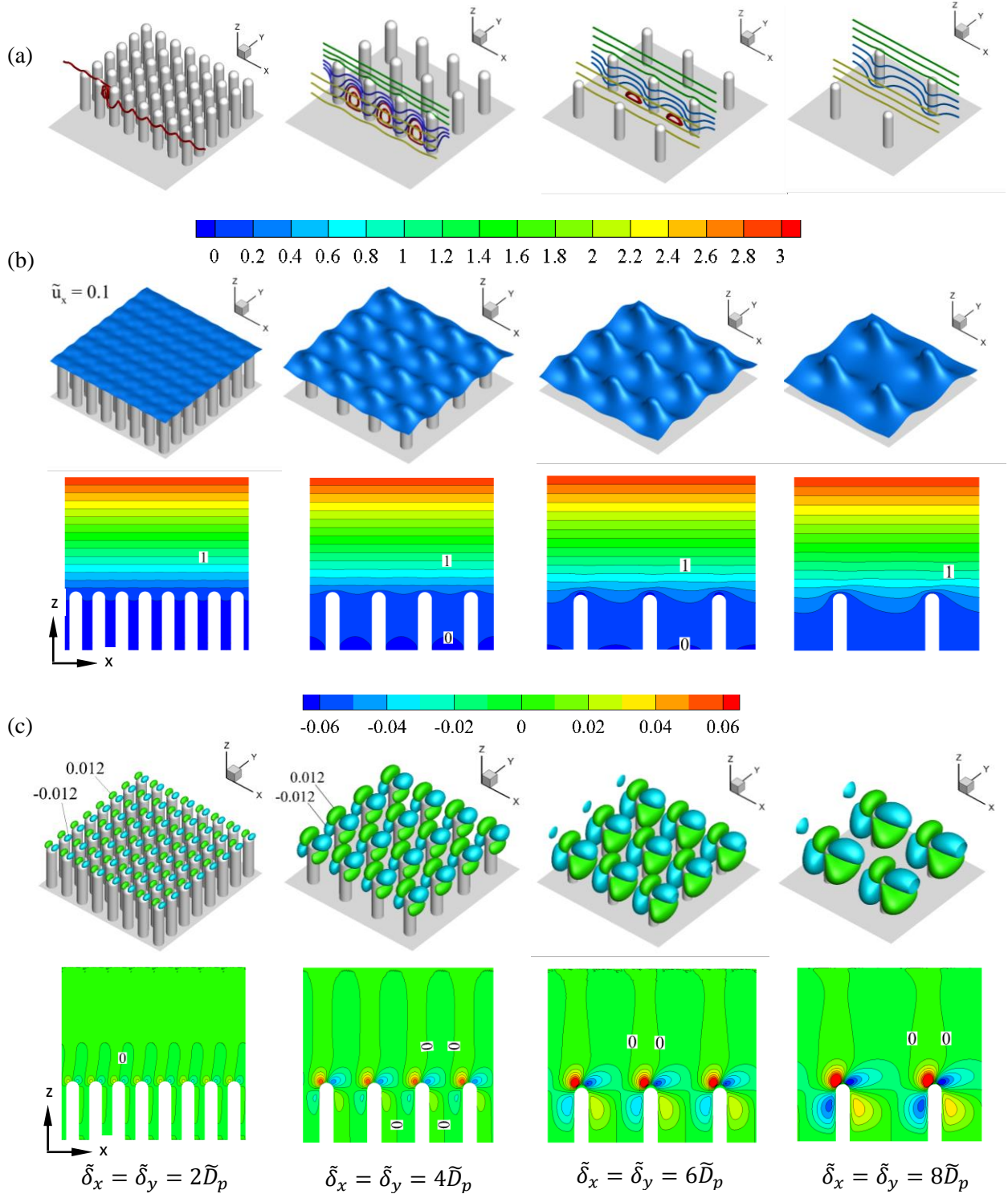

FIG. 5. Flow patterns for different streamwise and spanwise distances between micro pillars with $\tilde{\delta}_x = \tilde{\delta}_y$. (a) Typical streamlines, (b) streamwise velocity ($\tilde{u}_x$), and (c) vertical velocity ($\tilde{u}_z$).

Figure 6 shows the patterns of vertical gradient of streamwise velocity ($\partial \tilde{u}_x/\partial \tilde{z}$) for $\tilde{\delta}_x = \tilde{\delta}_y = 2\widetilde{D}_p$, $4\widetilde{D}_p$, $6\widetilde{D}_p$ and $8\widetilde{D}_p$. For different distances, the 3D iso-surfaces shown in Fig. 6(a) demonstrate similar patterns, yet the size of the tubular structures increases with the increase in $\tilde{\delta}_x$ and $\tilde{\delta}_y$. At the same time, the magnitude of $\partial \tilde{u}_x/\partial \tilde{z}$ around pillar tips increases. This is because the increase in $\tilde{\delta}_x$ and $\tilde{\delta}_y$ leads to the increase in the magnitudes of $\tilde{u}_x$ and $\tilde{u}_z$ around the pillar tips (Fig. 5), thereby causing the increase in $\partial \tilde{u}_x/\partial \tilde{z}$. For smaller $\tilde{\delta}_x$ and $\tilde{\delta}_y$, the smaller spaces among micro pillars restricts the momentum transport from the upper fluid to the fluid within the spaces, which causes a larger magnitude of $\partial \tilde{u}_x/\partial \tilde{z}$ in the region above pillar tips ($\tilde{z} > 1$) and a smaller magnitude below pillar tips ($\tilde{z} < 1$), as shown by the iso-contours on the streamwise plane through pillar centers in Fig. 6(b). With the increase in $\tilde{\delta}_x$ and $\tilde{\delta}_y$, the magnitude of $\partial \tilde{u}_x/\partial \tilde{z}$ decreases in the upper region, and increases in the spaces among micro pillars. Figure 6(c) shows the iso-contours of $\partial \tilde{u}_x/\partial \tilde{z}$ on the horizontal plane at the pillar height ($\tilde{z} = 1$), indicating that the mean value of $\partial \tilde{u}_x/\partial \tilde{z}$ in the areas between the streamwise pillar rows decreases as $\tilde{\delta}_x$ and $\tilde{\delta}_y$ increase from $2\widetilde{D}_p$ to $8\widetilde{D}_p$, yet the value of $\partial \tilde{u}_x/\partial \tilde{z}$ increases on the surface of pillar tips. These are associated with the strengthened interaction between the fluid and pillar tips and the enhanced momentum transport from the upper fluid to the lower fluid. On the bottom plane, as shown in Fig. 6(d), the mean value of $\partial \tilde{u}_x/\partial \tilde{z}$ increases with the increase in $\tilde{\delta}_x$ and $\tilde{\delta}_y$, due to the enhanced momentum transport to the lower region. When $\tilde{\delta}_x$ and $\tilde{\delta}_y$ further increase, the magnitude of $\partial \tilde{u}_x/\partial \tilde{z}$ will gradually approach the value of the flow confined by two infinitely large smooth planes, which is equal to 1.

In Fig. 7, the horizontally averaged quantities are compared for different pillar distances with $\tilde{\delta}_x = \tilde{\delta}_y$. The horizontally averaged quantity is defined as,

$$\langle \tilde{\beta} \rangle(\tilde{z}) \equiv \frac{1}{A_f} \int_{A_f} \tilde{\beta}(\tilde{x}, \tilde{y}, \tilde{z}) d\tilde{x} d\tilde{y} \qquad (5)$$

where $\tilde{\beta}(\tilde{x}, \tilde{y}, \tilde{z})$ is the quantity of interest, and $A_f$ is the area of horizontal plane occupied by fluid. $\langle \tilde{\beta} \rangle$ is the function of vertical coordinate ($\tilde{z}$). Since $\tilde{\delta}_x$ is equal to $\tilde{\delta}_y$, $\tilde{\delta}_{xy}$ is used to represent the identical $\tilde{\delta}_x$ and $\tilde{\delta}_y$ in the figure. Figure 7(a) shows the horizontally averaged streamwise velocity ($\langle \tilde{u}_x \rangle$). Above the pillar tips ($\tilde{z} > 1$), $\langle \tilde{u}_x \rangle$ changes linearly with $\tilde{z}$ for all $\tilde{\delta}_{xy}$, i.e., $\langle \partial \tilde{u}_x / \partial \tilde{z} \rangle$ is constant. Below the pillar tips ($\tilde{z} < 1$), $\tilde{u}_x$ gradually decreases to 0 toward the bottom plane. The variation of $\tilde{\delta}_{xy}$ causes apparent deviation in the curves. For $\tilde{\delta}_{xy} = 2\widetilde{D}_p$, $\tilde{u}_x$ decreases linearly from 3 at the top plane to a small value close to 0 at the pillar tips. With the increase in $\tilde{\delta}_{xy}$, the momentum transport in vertical direction is enhanced, which increase the value of $\langle \tilde{u}_x \rangle$ both above and below the pillar tips. The curve gradually approaches the straight line which corresponds to the flow over smooth bottom wall. Figure 7(b) shows the horizontally averaged vertical gradient of streamwise velocity ($\langle \partial \tilde{u}_x / \partial \tilde{z} \rangle$). The variation of $\langle \partial \tilde{u}_x / \partial \tilde{z} \rangle$ is consistent with that of $\langle \tilde{u}_x \rangle$ shown in Fig. 7(a). Above the pillar tips, $\langle \partial \tilde{u}_x / \partial \tilde{z} \rangle$ is constant, and the value is larger for smaller $\tilde{\delta}_{xy}$. Below the pillar tips, $\langle \partial \tilde{u}_x / \partial \tilde{z} \rangle$ is reduced due to the reduction of $\tilde{u}_x$, and gradually decrease towards the bottom plane. Contrary to the trend above the pillar tips, $\langle \partial \tilde{u}_x / \partial \tilde{z} \rangle$ is larger for larger $\tilde{\delta}_{xy}$ in the spaces ($\tilde{z} < 1$) due to the enhanced momentum transport. Both above and below the pillar tips, $\langle \partial \tilde{u}_x / \partial \tilde{z} \rangle$ approaches to 1 with the increase in $\tilde{\delta}_{xy}$.

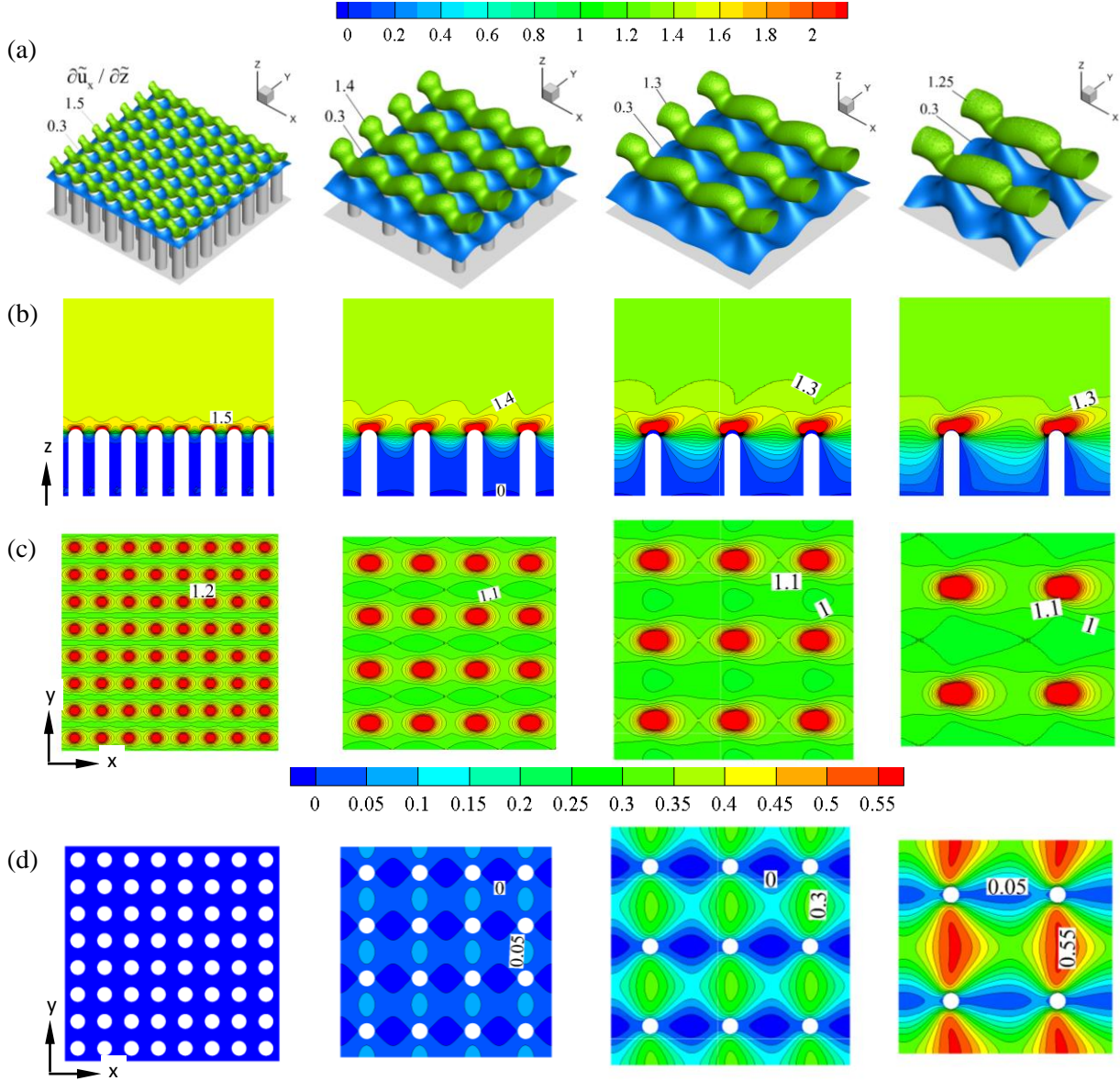

FIG. 6. Patterns of vertical gradient of streamwise velocity ($\partial \tilde{u}_x / \partial \tilde{z}$) for different streamwise and spanwise distances between micro pillars with $\tilde{\delta}_x = \tilde{\delta}_y$. (a) iso-surfaces, (b) iso-contours on a streamwise plane through pillar centers, (c) iso-contours on a horizon planes at pillar tips ($\tilde{z} = 1$), (d) iso-contours on the bottom plane ($\tilde{z} = 0$).

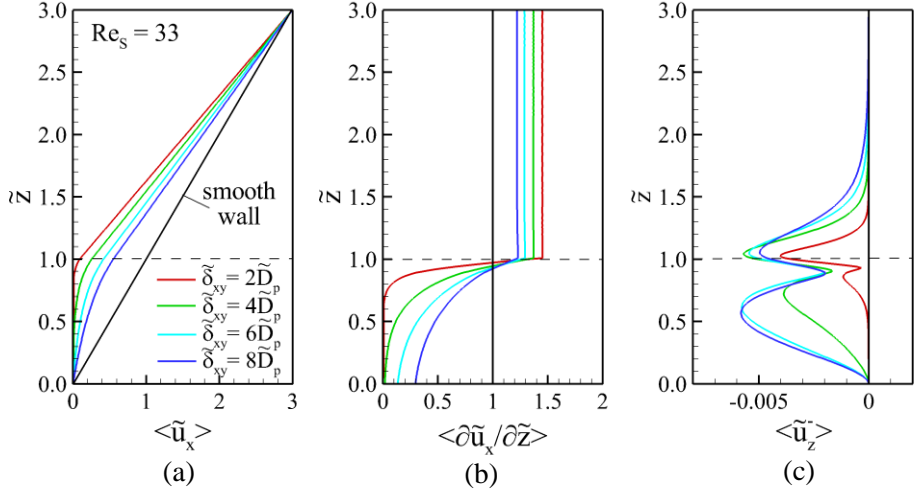

FIG. 7. Profiles of horizontally averaged quantities at different spacings between pillars for $\tilde{\delta}_x = \tilde{\delta}_y$. (a) Axial velocity $\langle \tilde{u}_x \rangle$, (b) vertical gradient of axial velocity $\langle \partial \tilde{u}_x / \partial \tilde{z} \rangle$, and (c) effective downward velocity $\langle \tilde{u}_z^- \rangle$.

The horizontally averaged vertical velocity $\langle \tilde{u}_z \rangle$ represents the net volume flux of fluid through a horizontal plane. As a result of mass conservation, $\langle \tilde{u}_z \rangle$ is zero for each individual structure unit in the present problem. To quantify the wall-normal advective transfer capability, we define the effective upward and downward velocity as

$$\tilde{u}_z^+ \equiv \begin{cases} \tilde{u}_z, & \text{if } \tilde{u}_z > 0 \\ 0, & \text{if } \tilde{u}_z < 0 \end{cases} \quad \text{and} \quad \tilde{u}_z^- \equiv \begin{cases} 0, & \text{if } \tilde{u}_z > 0 \\ \tilde{u}_z, & \text{if } \tilde{u}_z < 0 \end{cases} \qquad (6)$$

In the general heat and mass transfer problems, we work more often with $\tilde{u}_z^-$ than $\tilde{u}_z^+$. Figure 7(c) shows the variation of the effective downward velocity $\langle \tilde{u}_z^- \rangle$. Each curve has two peaks, corresponding to the upward and downward movement of the overhead flow around the pillar tips and the recirculation of micro eddies in the gaps between neighboring pillars. The absolute values of the peaks ($|\langle \tilde{u}_z^- \rangle|$) roughly reflect the advective transport capability in wall-normal direction. For $\tilde{\delta}_{xy} = 2\tilde{D}_p$, the small spaces among the pillars suppresses both the oscillation of the overhead flow around pillar tips and the development of micro eddies, and the eddies are close to the pillar tips, as shown in Fig. 5(a). Therefore, the values of $|\langle \tilde{u}_z^- \rangle|$ at both peaks are smaller for $\tilde{\delta}_{xy} = 2\tilde{D}_p$. When $\tilde{\delta}_{xy}$ increases from $2\tilde{D}_p$ to $6\tilde{D}_p$, the values of $|\langle \tilde{u}_z^- \rangle|$ at both peaks increase and the lower peak moves towards the bottom plane. The variation is consistent with the patterns of $\tilde{u}_z$ for different $\tilde{\delta}_{xy}$ shown in Fig. 5(c). When $\tilde{\delta}_{xy}$ increases from $6\tilde{D}_p$ to $8\tilde{D}_p$, the upper peak value of $|\langle \tilde{u}_z^- \rangle|$ decreases, and the lower one remains roughly unchanged. A turning point is expected for the lower peak where the peak value of $|\langle \tilde{u}_z^- \rangle|$ starts to decrease as $\tilde{\delta}_{xy}$ further increases. This is because the increase in $\tilde{\delta}_{xy}$ leads to the decrease in the number density of micro pillars, which dilutes the disturbance in the average sense. Further increase in $\tilde{\delta}_{xy}$ will make the peak values of $|\langle \tilde{u}_z^- \rangle|$ keep decreasing and approach zero gradually.

The variations of streamwise and spanwise pillar distances, $\tilde{\delta}_x$ and $\tilde{\delta}_y$ appear to have different effects on the characteristics of fluid flow. In the following discussion, two sets of special cases with either a small $\tilde{\delta}_x$ or a small $\tilde{\delta}_y$ are analyzed with the purpose to acquire a better understanding of the influence of $\tilde{\delta}_x$ and $\tilde{\delta}_y$.

Figure 8 shows the patterns of fluid flow with $\tilde{\delta}_x = 6\widetilde{D}_p$ and $\tilde{\delta}_y = 2\widetilde{D}_p$. The 2D iso-contours are plotted on a streamwise plane crossing the pillar centers. Fig. 8(a) shows the streamlines in the gaps between streamwise neighboring pillars (upper panel) and between the streamwise pillar rows (lower panel). The most prominent feature of this flow is the formation of a continuous tubular recirculating eddy, extending between the spanwise rows of micro pillars. In the flow with $\tilde{\delta}_x = \tilde{\delta}_y = 4\widetilde{D}_p$ shown in Figs. 2 and 3, the micro eddies in the gaps between the streamwise neighboring pillars are separated by the streamwise flow between the streamwise pillar rows. When the spanwise distance $\tilde{\delta}_y$ is decreased to $2\widetilde{D}_p$, the fluid between streamwise pillar rows gets involved in the recirculating motion of the eddies in the gaps, forming the continuous tubular recirculating eddies as shown in Fig. 8(a). At the same time, the streamwise momentum of the fluid within the spaces is reduced. Fig. 8(b) shows the 3D iso-surfaces (upper) and 2D iso-contours (lower) of the streamwise velocity $\tilde{u}_x$. The comparison of the patterns with the case of $\tilde{\delta}_x = \tilde{\delta}_y = 4\widetilde{D}_p$ (Fig. 2(b)) suggests that the momentum transport from the upper fluid to the fluid among micro pillars is significantly reduced by the tubular eddies. Fig. 8(c) shows the patterns of $\tilde{u}_z$ associated with the up-and-down movement of overhead flow around the pillar tips and the flow recirculation in the gaps below pillar tips. Due to the small spanwise distance ($\tilde{\delta}_y$), the variations of flow quantities, such as $\tilde{u}_x$ and $\tilde{u}_z$, are reduced in the spanwise direction, and the patterns demonstrate more 2D characteristics, as functions of $\tilde{x}$ and $\tilde{z}$.

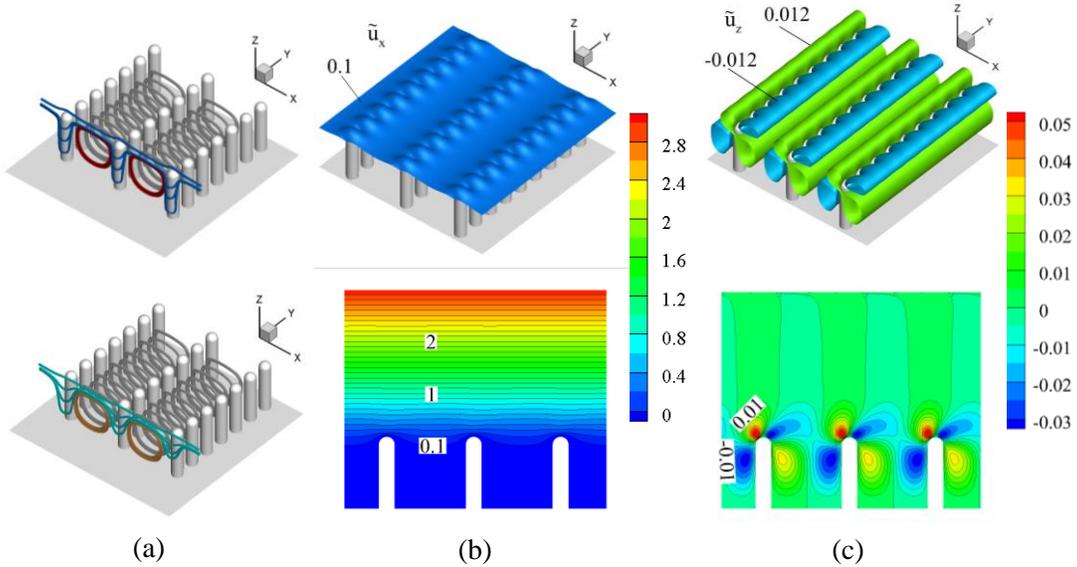

(a) (b) (c)

FIG. 8. Patterns of axial and vertical velocities for $\tilde{\delta}_x = 6\widetilde{D}_p$, $\tilde{\delta}_y = 2\widetilde{D}_p$. (a) 3D streamlines around pillars (Upper: streamlines between streamwise neighboring pillars, lower: streamlines between streamwise pillar rows), (b) axial velocity $\tilde{u}_x$, and (c) vertical velocity $\tilde{u}_z$. The iso-contours are on a streamwise plane through pillar centers.

Figure 9 shows the patterns of $\partial \tilde{u}_x / \partial \tilde{z}$ with $\tilde{\delta}_x = 6\widetilde{D}_p$ and $\tilde{\delta}_y = 2\widetilde{D}_p$. Similar to the flow with $\tilde{\delta}_x = \tilde{\delta}_y = 4\widetilde{D}_p$ shown in Fig. 3, a region with larger $\partial \tilde{u}_x / \partial \tilde{z}$ is formed around the pillar tips due to the sharp turn of the overhead flow. Because of the small $\tilde{\delta}_y$ and larger $\tilde{\delta}_x$, the regions of larger $\partial \tilde{u}_x / \partial \tilde{z}$ around the pillar tips connect with each other in the spanwise direction, and forms a tubular structure along the pillar tips. Figure 9(b) shows the distribution of $\partial \tilde{u}_x / \partial \tilde{z}$ on the bottom plane among the pillars. The continuous tubular recirculating eddies shown in Fig. 8(a) create a banded area with negative $\partial \tilde{u}_x/\partial \tilde{z}$ between every pair of spanwise pillar rows. Between spanwise neighboring pillars, $\partial \tilde{u}_x/\partial \tilde{z}$ is positive on the bottom plane.

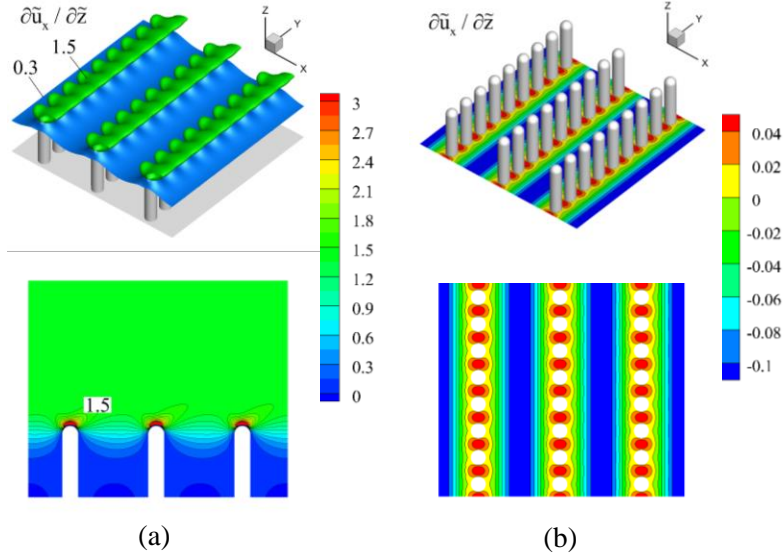

(a) (b)

FIG. 9. Patterns of vertical gradient of axial velocity ($\partial \tilde{u}_x/\partial \tilde{z}$) for $\tilde{\delta}_x = 6\tilde{D}_p$, $\tilde{\delta}_y = 2\tilde{D}_p$. (a) Entire regime, and (b) at bottom surface.

For the fluid flows with a small spanwise pillar distance ($\tilde{\delta}_y$), the flow characteristics and transport capabilities are also influenced by the streamwise pillar distance ($\tilde{\delta}_x$). In Fig. 10, the horizontally averaged streamwise velocity $\langle \tilde{u}_x \rangle$, vertical gradient of streamwise velocity $\langle \partial \tilde{u}_x/\partial \tilde{z} \rangle$ and vertical velocity $\langle \tilde{u}_z \rangle$ are compared for different $\tilde{\delta}_x$. $\tilde{\delta}_y$ is fixed at $2\tilde{D}_p$, and $\tilde{\delta}_x$ ranges from $2\tilde{D}_p$ to $8\tilde{D}_p$. As shown in Fig. 10(a), the curves of $\langle \tilde{u}_x \rangle$ are close to each other. Above the pillar tips ($\tilde{z} > 1$), $\langle \tilde{u}_x \rangle$ decreases linearly from 3 at top plane to a value close to 0 at pillar tips. Within the spaces among pillars ($\tilde{z} < 1$), $\langle \tilde{u}_x \rangle$ is close to zero. This variation suggests that the momentum transport from the upper fluid to the fluid within the spaces among micro pillars is significantly blocked by the tubular eddies. This conclusion can also be made from the variation of $\langle \partial \tilde{u}_x/\partial \tilde{z} \rangle$ shown in Fig. 10(b). Above the pillar tips, the values of $\langle \partial \tilde{u}_x/\partial \tilde{z} \rangle$ for different $\tilde{\delta}_x$ are constant and close to 1.5 which corresponds the flow over a smooth plane at pillar tips. Within the spaces among pillars, the value of $\langle \partial \tilde{u}_x/\partial \tilde{z} \rangle$ decreases sharply to a small value. The curves show apparent deviation for different $\tilde{\delta}_x$ due to the different sizes of recirculating eddies. The value of $\langle \partial \tilde{u}_x/\partial \tilde{z} \rangle$ decreases smoother for larger $\tilde{\delta}_x$, indicating a relatively larger momentum flux from the upper fluid. The variation of effective downward velocity ($\langle \tilde{u}_z^- \rangle$) is shown in Fig. 10(c). Two peaks with negative values of $\langle \tilde{u}_z^- \rangle$ can be observed above and below the pillar tips, corresponding to the up and down movement of the overhead flow and the flow recirculation within the gaps. Unlike $\langle \tilde{u}_x \rangle$ and $\langle \partial \tilde{u}_x/\partial \tilde{z} \rangle$, the curves of $\langle \tilde{u}_z^- \rangle$ show apparent deviation for different $\tilde{\delta}_x$. When $\tilde{\delta}_x$ increases from $2\tilde{D}_p$ to $6\tilde{D}_p$, $|\langle \tilde{u}_z^- \rangle|$ at both peaks increases significantly due to the enhanced momentum transport. This trend is destroyed by the dilution effect of decreased number density of micro pillars. After a certain critical value of $\tilde{\delta}_x$, the values of $|\langle \tilde{u}_z^- \rangle|$ at the two peaks will decrease with the increase in $\tilde{\delta}_x$ and gradually approach 0. In the figure, when $\tilde{\delta}_x$ increases from $6\tilde{D}_p$ to $8\tilde{D}_p$, the value of $|\langle \tilde{u}_z^- \rangle|$ at the lower peak keeps increasing, yet the value at the upper peak changes very little. This suggests that the critical value of $\tilde{\delta}_x$ for the upper peak has been reached. Further increase in $\tilde{\delta}_x$ will leads to the decrease in $|\langle \tilde{u}_z^- \rangle|$ at the upper peak and brings up the critical $\tilde{\delta}_x$ for the lower peak.

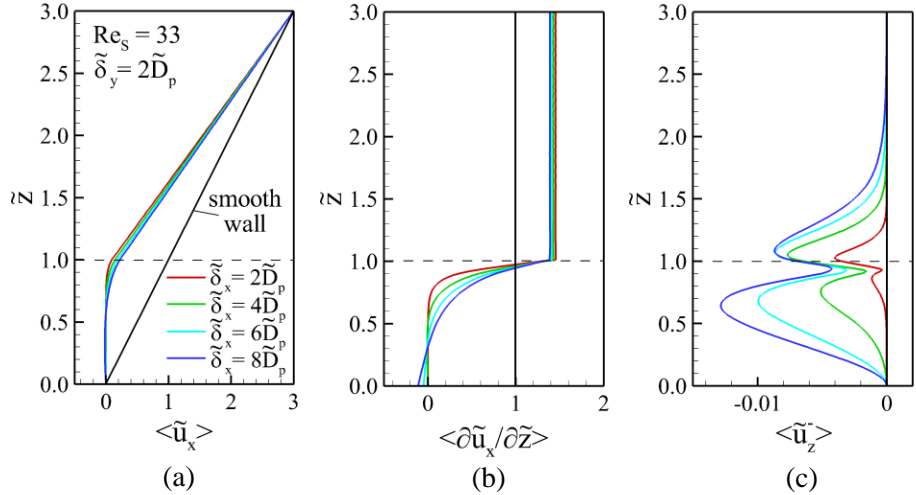

(a) (b) (c)

FIG. 10. Profiles of horizontally averaged quantities at different streamwise spacings between pillars for $\tilde{\delta}_y = 2\tilde{D}_p$. (a) Axial velocity $\langle \tilde{u}_x \rangle$, (b) vertical gradient of axial velocity $\langle \partial \tilde{u}_x/\partial \tilde{z} \rangle$, and (c) effective downward velocity $\langle \tilde{u}_z^- \rangle$.

Another special set of cases with a small streamwise distance ($\tilde{\delta}_x = 2\tilde{D}_p$) is also considered in this study. Figure 11 shows the flow patterns with $\tilde{\delta}_x = 2\tilde{D}_p$ and $\tilde{\delta}_y = 6\tilde{D}_p$. The 2D iso-contours are plotted on a spanwise plane crossing the pillar centers. Because of the small size of the gaps between streamwise neighboring pillars and the large scale of the streamwise flow between the streamwise pillar rows, no recirculating eddy is generated, as shown by the typical streamlines in Fig. 11(a). The patterns of streamwise velocity ($\tilde{u}_x$) shown in Fig. 11(b) indicate that the magnitude of $\tilde{u}_x$ is much larger than that in the flow with $\tilde{\delta}_x = 6\tilde{D}_p$ and $\tilde{\delta}_y = 2\tilde{D}_p$. This implies that the momentum transport from the upper fluid to the fluid in the spaces among micro pillars is significantly enhanced. Due to the absence of the micro eddies, the disturbance of $\tilde{u}_z$ mainly appears around the pillar tips, as shown in Fig. 11(c). The iso-surfaces with positive and negative values of $\tilde{u}_z$ are staggered to form a streamwise columnar area along the pillar tips. Because of the small $\tilde{\delta}_x$, the variation of flow quantities is reduced in the streamwise direction, and the flow patterns demonstrate more 2D characteristics, as functions of $\tilde{y}$ and $\tilde{z}$.

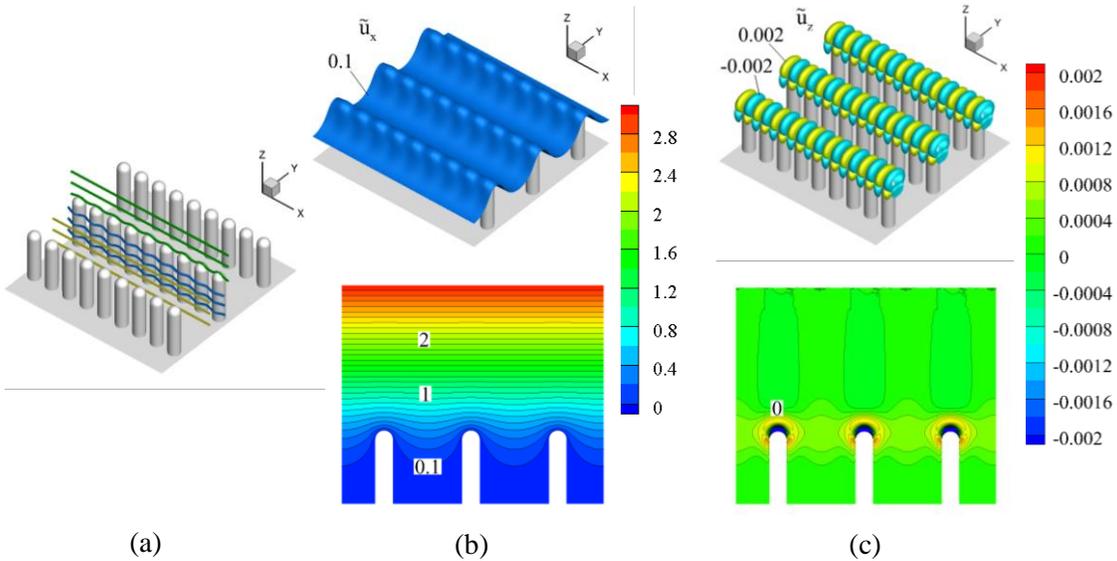

(a)            (b)            (c)

FIG. 11. Patterns of axial and vertical velocities for $\tilde{\delta}_x = 2\tilde{D}_p$, $\tilde{\delta}_y = 6\tilde{D}_p$. (a) 3D streamlines around micro pillars, (b) axial velocity $\tilde{u}_x$, and (c) vertical velocity $\tilde{u}_z$. The iso-contours are on a spanwise plane through pillar centers.

The patterns of the vertical gradient of streamwise velocity ($\partial \tilde{u}_x/\partial \tilde{z}$) are shown in Fig. 12. Because of the small $\tilde{\delta}_x$, the regions with larger $\partial \tilde{u}_x/\partial \tilde{z}$ around the pillar tips connect with each other in the streamwise direction and form a tubular region along the pillar tips, as shown in Fig. 12(a). The enhanced momentum flux increases the magnitude of $\partial \tilde{u}_x/\partial \tilde{z}$ in the spaces between streamwise pillar rows, and creates a banded area with positive $\partial \tilde{u}_x/\partial \tilde{z}$ between the spanwise pillar rows on the bottom plane, as shown in Fig. 12(b). In the gaps between the streamwise neighboring pillars, $\partial \tilde{u}_x/\partial \tilde{z}$ is close to 0 because the fluid is almost stagnant in the gaps.

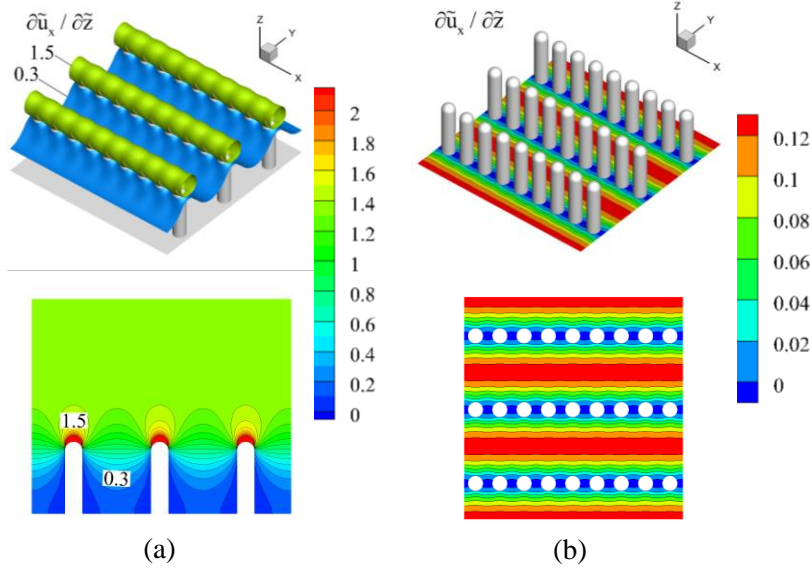

FIG. 12. Patterns of vertical gradient of axial velocity ($\partial \tilde{u}_x/\partial \tilde{z}$) for $\tilde{\delta}_x = 2\tilde{D}_p$, $\delta_y = 6\tilde{D}_p$.
(a) Entire regime, and (b) at the bottom plane.

The profiles of the horizontally averaged streamwise velocity ($\langle \tilde{u}_x \rangle$), vertical gradient of streamwise velocity ($\langle \partial \tilde{u}_x/\partial \tilde{z} \rangle$) and vertical velocity ($\langle \tilde{u}_z \rangle$) for different spanwise distance ($\tilde{\delta}_y$) are shown in Fig. 13. $\tilde{\delta}_x$ is fixed at $2\tilde{D}_p$, and $\tilde{\delta}_y$ ranges from $2\tilde{D}_p$ to $8\tilde{D}_p$. As shown in Fig. 13(a), for all $\tilde{\delta}_y$, $\langle \tilde{u}_x \rangle$ decreases linearly from 3 at the top plane to a small value close to 0 at the pillar height. Below the pillar tips, $\langle \tilde{u}_x \rangle$ decreases to 0 at the bottom plane. With the increase in $\tilde{\delta}_y$, the momentum transport from the upper fluid to the fluid within the spaces is enhanced, which increases the value of $\langle \tilde{u}_x \rangle$ at each vertical level above and below the pillar tips. Over the entire range, the curve of $\langle \tilde{u}_x \rangle$ gradually approaches the straight line which corresponds to the flow confined by smooth planes, as $\tilde{\delta}_y$ increases from $2\tilde{D}_p$ to $8\tilde{D}_p$. The curves of $\langle \partial \tilde{u}_x/\partial \tilde{z} \rangle$ shown in Fig. 13(b) are consistent with the curves of $\langle \tilde{u}_x \rangle$. Above the pillar tips ($\tilde{z} > 1$), $\langle \partial \tilde{u}_x/\partial \tilde{z} \rangle$ is constant for each $\tilde{\delta}_y$, and the value decreases with the increase in $\delta_y$. Below the pillar tips, however, the value of $\langle \partial \tilde{u}_x/\partial \tilde{z} \rangle$ increases with the increase in $\tilde{\delta}_y$ at every vertical level, due to the enhanced momentum transport. Figure 13(c) shows the profiles of the effective downward velocity $\langle \tilde{u}_z^- \rangle$. Compared with the upper peak around the pillar tips, the value of $|\langle \tilde{u}_z^- \rangle|$ at the lower peak is significantly reduced, due to the suppression of eddies in the gaps. With the increase in $\tilde{\delta}_y$, the increased streamwise velocity in the spaces between the streamwise pillar rows further suppresses the eddies in the gaps. The increased streamwise velocity also suppresses the oscillation of overhead flow around the pillar tips. At the same time, the dilution effect of the decreased number density of micro pillars takes effect. These two factors cause a decrease in $|\langle \partial \tilde{u}_x/\partial \tilde{z} \rangle|$ at both peaks. Compared with other flows, the flows with small $\tilde{\delta}_x$ are relatively simpler.

In the above discussion, the basic flow patterns for different streamwise and spanwise pillar distances for $Re = 33$ were analyzed. The cases with the same pillar arrangements for $Re = 1$ are also considered in this study. The flows for two Reynolds numbers have similar spatial distribution of velocity components. The difference lies mainly in the influence of fluid inertia on flow characteristics. When $Re = 1$, the flow patterns on the windward and leeward sides are symmetric about the pillar center, and the flow advection only includes the recirculation of micro eddies and the oscillation of the fluid surrounding the micro pillars, which are referred to as local advection. When $Re = 33$, the strengthened fluid inertia breaks the symmetry, and creates a spiral long-range advection. In comparison, the flow characteristics are $Re = 1$ are relatively simpler, so this paper will not go into details.

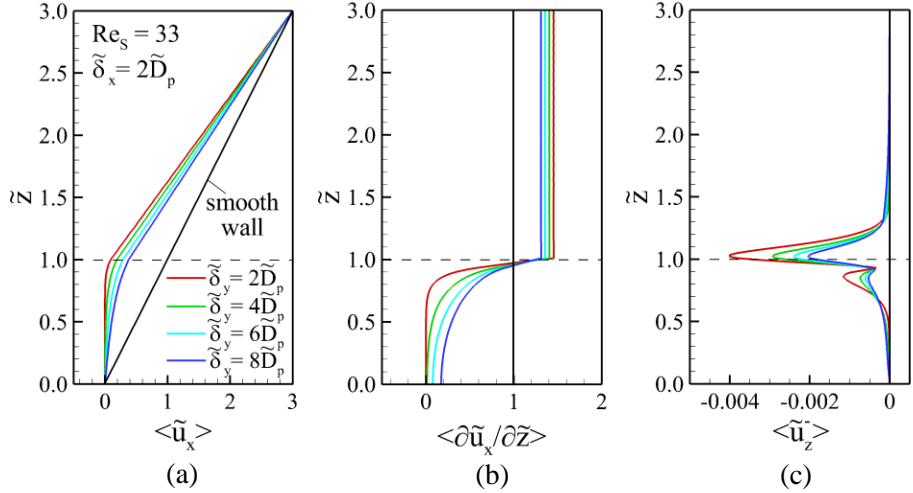

FIG. 13. Profiles of horizontally averaged quantities at different spanwise spacings between pillars for $\tilde{\delta}_x = 2\tilde{D}_p$. (a) Axial velocity $\langle \tilde{u}_x \rangle$, (b) vertical gradient of axial velocity $\langle \partial \tilde{u}_x / \partial \tilde{z} \rangle$, and (c) effective downward velocity $\langle \tilde{u}_z^- \rangle$.

### *Decomposed surface friction on structured wall*

On micro-structured surfaces, the total surface drag ($F_{tot}$) over each structure unit includes three components: the reaction force of a single pillar due to the flow shear stress at pillar surface ($F_{p,s}$), the reaction force of a single pillar due to the flow pressure at pillar surfaces ($F_{p,p}$), and the reaction force of the bottom plane due to the flow shear stress at bottom surface ($F_{b,s}$). The forces are obtained by integrating the streamwise component of the shear stress and pressure over the surface area of micro pillars or bottom plane. $F_{p,s}$ and $F_{p,p}$ plays a dominant role in determining $F_{tot}$, and are also the important parameters that need special attention in the design of surface structures. In this study, the forces are nondimensionalized as,

$$\tilde{F} = \frac{F}{(\mu U_0/H)(\pi R_p^2)} \tag{7}$$

where $R_p$ is the radius of micro pillars. Figure 14 shows the 2D iso-contours of the reaction force of shear stress ($\tilde{F}_{p,s}$) and pressure ($\tilde{F}_{p,p}$) of a single pillar in the space of $\tilde{\delta}_x$ and $\tilde{\delta}_y$ for $Re = 1$ and 33, respectively. As shown in the figure, both $\tilde{F}_{p,s}$ and $\tilde{F}_{p,p}$ demonstrate similar dependences on $\tilde{\delta}_x$ and $\tilde{\delta}_y$ between $Re = 1$ and 33. The comparison shows that $\tilde{F}_{p,s}$ is smaller and $\tilde{F}_{p,p}$ is larger for $Re = 33$ at each point of $\tilde{\delta}_x$ and $\tilde{\delta}_y$. This difference is caused by the effect of fluid inertia at larger Reynolds numbers, and has been discussed in detail in [17]. For densely arranged micro pillars, the shielding effect, which is associated with both $\tilde{\delta}_x$ and $\tilde{\delta}_y$, reduces $\tilde{F}_{p,s}$ and $\tilde{F}_{p,p}$. As $\tilde{\delta}_x$ and $\tilde{\delta}_y$ increase, the shielding effect is reduced and the pillar surface area exposed to the upper bulk flow increases. As a result, both $\tilde{F}_{p,s}$ and $\tilde{F}_{p,p}$ increase with $\tilde{\delta}_x$ and $\tilde{\delta}_y$, as shown in the figure.

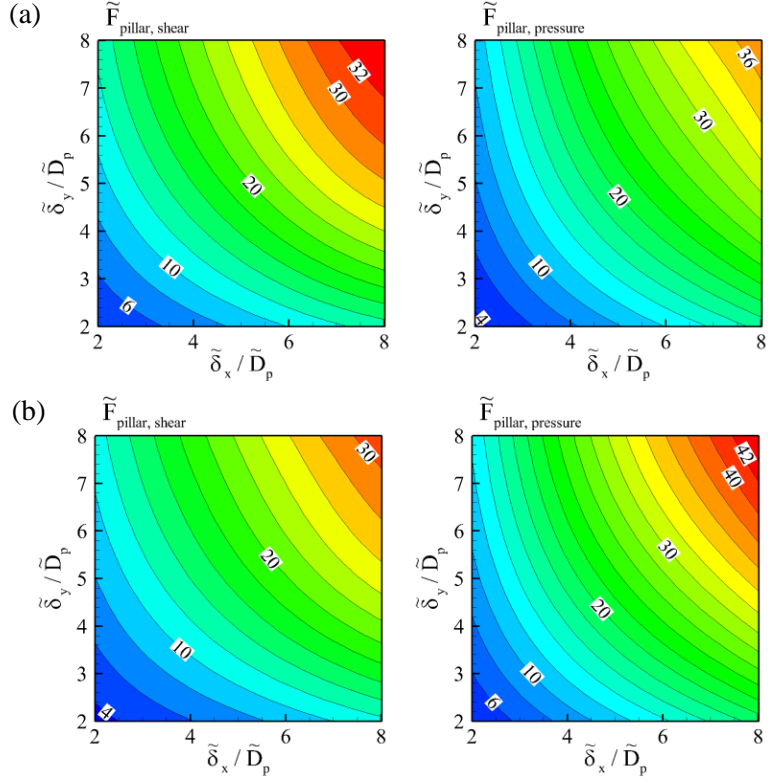

FIG. 14. Distribution of shear (left) and pressure (right) forces exerted on each pillar in the space of $\tilde{\delta}_x$ and $\tilde{\delta}_y$. (a) $Re = 1$, and (b) $Re = 33$.

For different pillar arrangements, the surface friction can be compared over a unit horizontal area. The equivalent shear stresses on the structured surfaces for the total friction, reaction force of micro pillars due to flow shear and flow pressure, and reaction force of bottom surface due to flow shear are defined as,

$$\tilde{T} = \frac{\pi R_p^2}{A} \tilde{F} \tag{8}$$

where $\tilde{F}$ is the total surface drag or the components of reaction forces of a single structure unit, and $A$ is the horizontal area of a single structure unit. The equivalent shear stress of total friction ($\tilde{T}_{tot}$) is written as,

$$\tilde{T}_{tot} = \tilde{T}_{p,s} + \tilde{T}_{p,p} + \tilde{T}_{b,s} \tag{9}$$

where $\tilde{T}_{p,s}$, $\tilde{T}_{p,p}$, and $\tilde{T}_{b,s}$ are the equivalent shear stresses of $\tilde{F}_{p,s}$, $\tilde{F}_{p,p}$, and $\tilde{F}_{b,s}$.

Figure 15 and 16 shows the iso-contours of $\tilde{T}_{tot}$, $\tilde{T}_{p,s}$, $\tilde{T}_{p,p}$ and $\tilde{T}_{b,s}$ in the space of $\tilde{\delta}_x$ and $\tilde{\delta}_y$ for $Re = 1$ and 33, respectively. Compared with the shear stress force ($\tilde{F}_{p,s}$) and pressure force ($\tilde{F}_{p,p}$) shown in Fig. 14, $\tilde{T}_{tot}$, $\tilde{T}_{p,s}$, $\tilde{T}_{p,p}$ and $\tilde{T}_{b,s}$ demonstrate more complex dependences on $\tilde{\delta}_x$ and $\tilde{\delta}_y$. These complex dependences are mainly composed of two factors, the dilution effect of the decreasing number density of micro pillars and the multi-faceted effects of micro eddies in the gaps between the streamwise neighboring pillars. The distributions of these equivalent shear stresses show similar patterns for $Re = 1$ and 33. The difference is associated with the fluid inertia which has been discussed in [17]. Here we focus on the case for $Re = 1$ to analyze the pillar arrangement effect.

The pattern of the equivalent shear stress of total friction ($\tilde{T}_{tot}$) is shown in Fig. 15(a), which exhibits a relatively simple relationship with $\tilde{\delta}_x$ and $\tilde{\delta}_y$. When the micro pillars are closely embedded, that is, $\tilde{\delta}_x$ and $\tilde{\delta}_y$ are small, $\tilde{T}_{tot}$ is large. One extreme case is that all the spaces among the micro pillars are filled, then $\tilde{T}_{tot}$ is,

$$\tilde{T}|_{\tilde{\delta}_x,\tilde{\delta}_y \to 0} = \tilde{U}_0/(\tilde{H} - \tilde{l}_p) = 1.5 \tag{10}$$

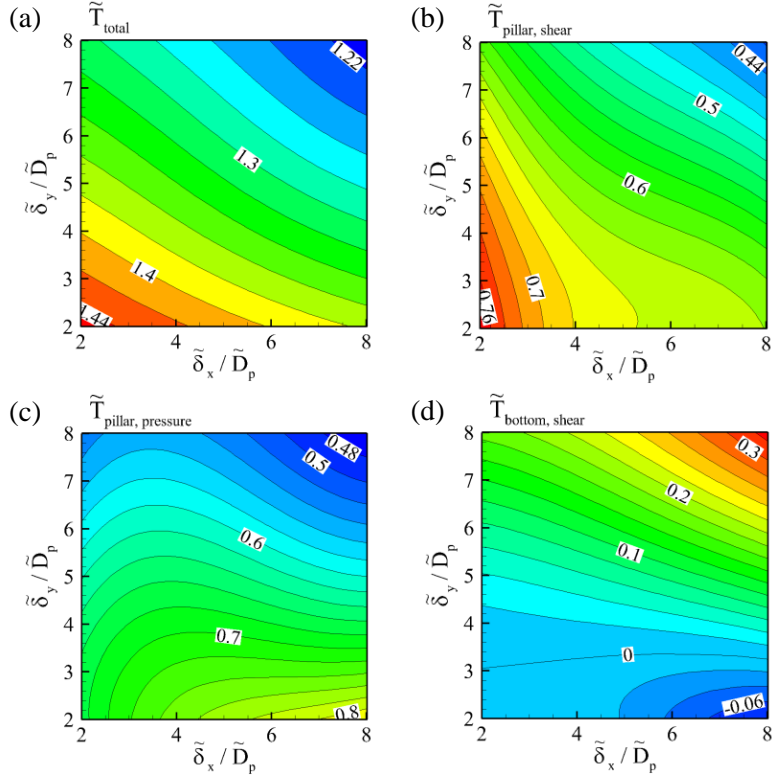

FIG. 15. Distribution of (a) equilent shear stress of total friction ($\tilde{T}_{tot,}$), (b) equivalent shear stress of reaction force due to flow shear at pillar surface ($\tilde{T}_{p,s}$), (c) equivalent shear stress of reaction force due to flow pressure at pillar surface ($\tilde{T}_{p,p}$), and (d) equivalent shear stress of reaction force due to flow shear at bottom surface ($\tilde{T}_{b,s}$), in the space of $\tilde{\delta}_x$ and $\tilde{\delta}_y$ at $Re = 1$.

In the nondimensional form, the viscosity is not included in Eqn. (10). The result of our numerical simulation is $\tilde{T}_{tot} = 1.45$ when $\tilde{\delta}_x = \tilde{\delta}_y = 2\tilde{D}_p$. The numerical result is slightly less than 1.5 because the small spaces among the micro pillars reduces the friction. On the whole, $\tilde{T}_{tot}$ exhibits a simple monotonically decreasing dependence on $\tilde{\delta}_x$ and $\tilde{\delta}_y$. Any increase in $\tilde{\delta}_x$ or $\tilde{\delta}_y$ will lead to a decrease in $\tilde{T}_{tot}$. When $\tilde{\delta}_x$ and $\tilde{\delta}_y$ become infinitely large, the flow will reduce to that confined by two smooth planes, and the equivalent shear stress of total friction will be,

$$\tilde{T}|_{\tilde{\delta}_x,\tilde{\delta}_y \to \infty} = \tilde{U}_0/\tilde{H} = 1 \tag{11}$$

In the simulation, $\tilde{T}_{tot} = 1.21$ when $\tilde{\delta}_x = \tilde{\delta}_y = 8\tilde{D}_p$. Behind this simple relationship is the complex dependence of the three components on $\tilde{\delta}_x$ and $\tilde{\delta}_y$.

Figure 15(b) shows the iso-contours of the equivalent shear stress of the reaction force of a single pillar due to flow shear at pillar surfaces ($\tilde{T}_{p,s}$). Within the whole area, the distribution of $\tilde{T}_{p,s}$ shows a trend of decreasing with the increase in $\tilde{\delta}_x$ and $\tilde{\delta}_y$. It has been shown in Fig. 14 that the reaction force of a single pillar ($\tilde{F}_{p,s}$) increases with the increased $\tilde{\delta}_x$ and $\tilde{\delta}_y$. Therefore, the decreasing trend of $\tilde{T}_{p,s}$ is caused by the dilution effect of decreasing pillar number density. In the lower right corner of the figure, the distribution of $\tilde{T}_{p,s}$ shows an obvious distortion. This is because that in that area the spanwise distance $\tilde{\delta}_y$ is small, which facilitates the generation of the tubular eddies as shown in Fig. 8 and 9. The tubular eddies enhances the momentum transport from the upper fluid to the fluid within the spaces among micro pillars, thereby decreasing the reaction force due to flow shear and shifting the contour lines of larger values to the left. On the other hand, in the lower right corner $\tilde{\delta}_x$ is comparable to the pillar length and is in the favorable range for stimulating the recirculation of the eddies. Therefore, the iso-contours are distorted significantly in that area. With the further increase in $\tilde{\delta}_x$, the effects of micro eddies will be diluted by the decreasing number density of micro pillars. When $\tilde{\delta}_x$ and $\tilde{\delta}_y$ become infinitely large, $\tilde{T}_{p,s}$ will decrease to zero.

Figure 15(c) shows the iso-contours of the equivalent shear stress of the reaction force of a single pillar due to flow pressure ($\tilde{T}_{p,p}$). In the upper right corner, the iso-contours exhibit the trend of decreasing with the increasing $\tilde{\delta}_x$ and $\tilde{\delta}_y$, which is caused by the dilution effect of pillar density. Yet in the area of smaller $\tilde{\delta}_x$ and $\tilde{\delta}_y$, the contour lines are strongly distorted. The maximum of $\tilde{T}_{p,p}$ is reached in the lower right corner where the tubular eddies are generated between the spanwise rows of micro pillars (Fig. 8 and 9), implying that the distortion of contour lines are associated with the micro eddies in the spaces among pillars. In the lower left corner, both $\tilde{\delta}_x$ and $\tilde{\delta}_y$ are small, a micro eddy with a smaller size is generated in each gap between each pair of streamwise neighboring pillars. As $\tilde{\delta}_x$ increases, the eddy size increases correspondingly. Due to the small spanwise distance ($\tilde{\delta}_y$), the discrete eddies in the gaps connect with each other in the spanwise direction in the process of size increasing of the eddies. The tubular eddies prevent the penetration of upper fluid into the spaces among micro pillars and make the overhead flow impact the pillar rows more strongly. Therefore, $\tilde{T}_{p,p}$ increases with $\tilde{\delta}_x$ for a smaller $\tilde{\delta}_y$, as shown in the figure. After a critical point, further increase in $\tilde{\delta}_x$ will reduce the strength of the eddies. At the same time, the dilution effect of decreasing pillar number density will become dominant gradually. As a result, $\tilde{T}_{p,p}$ will decrease with the increasing $\tilde{\delta}_x$ when $\tilde{\delta}_x$ goes beyond a certain critical value. In the lower right corner (larger $\tilde{\delta}_x$ and smaller $\tilde{\delta}_y$), the flow is characterized as a tubular eddy between each pair of spanwise pillar rows. As $\tilde{\delta}_y$ increases, the increased spaces between the spanwise neighboring pillars enhance the momentum transport from the upper fluid to the fluid in the spaces and gradually break the tubular eddies. Therefore, $\tilde{T}_{p,p}$ decreases with the increase in $\tilde{\delta}_y$. The dilution effect of decreasing pillar number density also contributes to this process. This trend can also be observed for smaller $\tilde{\delta}_x$, but it is mainly caused by the dilution effect. In the upper left corner (smaller $\tilde{\delta}_x$ and larger $\tilde{\delta}_y$), the pillars are close to each other in the streamwise direction. With the increase in $\tilde{\delta}_x$, the pillar surface area exposed to the overhead flow increases, which increases the impact of overhead flow on micro pillars, so $\tilde{T}_{p,p}$ increases with $\tilde{\delta}_x$ in the range of $\tilde{\delta}_x < \approx 3.5 \tilde{D}_p$. After a certain critical point at $\tilde{\delta}_x \approx 3.5 \tilde{D}_p$, the dilution of decreasing number density of micro pillars takes effect, and $\tilde{T}_{p,p}$ decreases with the increasing $\tilde{\delta}_x$.

Figure 15(d) shows the iso-contours of the equivalent shear stress of the reaction force of bottom plane due to flow shear ($\tilde{T}_{b,s}$). In the area of larger $\tilde{\delta}_y$, the increase in $\tilde{\delta}_x$ and $\tilde{\delta}_y$ enhances the momentum transport to the bottom plane, so $\tilde{T}_{b,s}$ increases with the increase in $\tilde{\delta}_x$ and $\tilde{\delta}_y$. When $\tilde{\delta}_x$ and $\tilde{\delta}_y$ are infinitely large, the flow will reduce to that over a smooth plane, and $\tilde{T}_{b,s}$ will be

$$\tilde{T}_{b,s}\big|_{\tilde{\delta}_x,\tilde{\delta}_y \to \infty} = \tilde{U}_0/\tilde{H} = 1 \tag{12}$$

On the other hand, when $\tilde{\delta}_x$ and $\tilde{\delta}_y$ are both very small, the fluid within the spaces among micro pillars will be stagnant, and

$$\tilde{T}_{b,s}\big|_{\tilde{\delta}_x,\tilde{\delta}_y \to 0} = 0 \tag{13}$$

As $\tilde{\delta}_x$ increases, the micro eddies gradually increase in size and makes the flow shear rate negative beneath the eddies on the bottom plane, so $\tilde{T}_{b,s}$ is negative in the lower region of the figure. The increasing $\tilde{\delta}_x$ strengthens the micro eddies, and increases the magnitude of negative $\tilde{T}_{b,s}$. In the lower right corner ($\tilde{\delta}_x$ is large and $\tilde{\delta}_y$ is small), the development of tubular eddies between the spanwise pillar rows creates the banded areas with negative flow shear rate beneath the eddies (Fig. 8 and 9), which makes $\tilde{T}_{b,s}$ reach the minimum. After that, the effect of decreasing pillar number density becomes dominant. Further increase in $\tilde{\delta}_x$ will lead to the increase in $\tilde{T}_{b,s}$.

Figure 16 shows the iso-contours of $\tilde{T}_{tot}$, $\tilde{T}_{p,s}$, $\tilde{T}_{p,p}$ and $\tilde{T}_{b,s}$ for $Re = 33$. The dependences of these quantities on $\tilde{\delta}_x$ and $\tilde{\delta}_y$ are similar to those for $Re = 1$. Due to the enhanced fluid inertia, $\tilde{T}_{p,s}$ is smaller and $\tilde{T}_{p,p}$ is larger at each point of $\tilde{\delta}_x$ and $\tilde{\delta}_y$ for $Re = 33$. The effect of fluid inertia on $\tilde{T}_{tot}$, $\tilde{T}_{p,s}$, $\tilde{T}_{p,p}$ and $\tilde{T}_{b,s}$ for the case of $\tilde{\delta}_x = \tilde{\delta}_y = 4\tilde{D}_p$ has been analyzed in [17]. The analysis also applies to other combinations of $\tilde{\delta}_x$ and $\tilde{\delta}_y$.

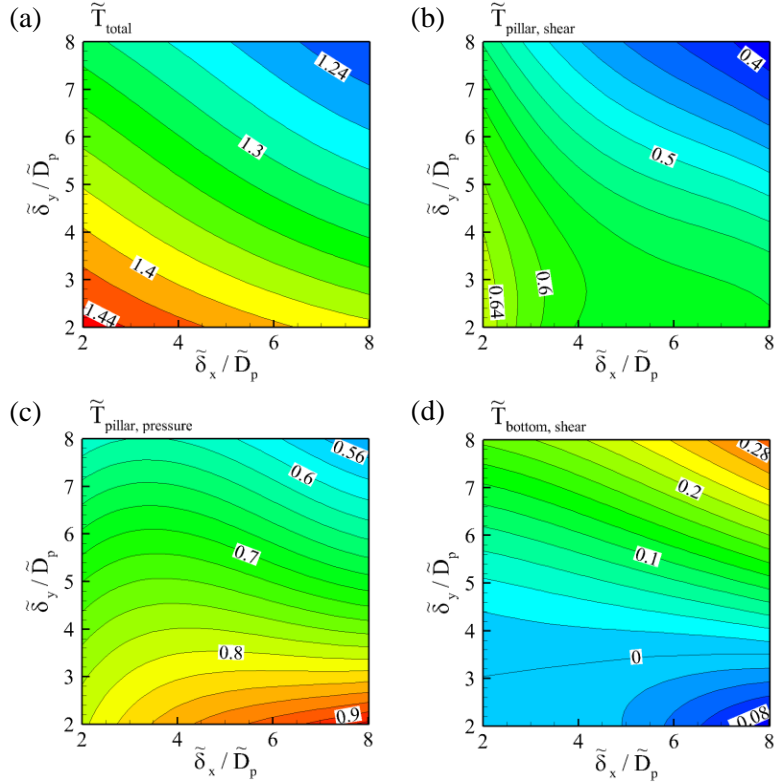

FIG. 16. Distribution of (a) equilent shear stress of total friction ($\tilde{T}_{tot}$,), (b) equivalent shear stress of reaction force due to flow shear at pillar surface ($\tilde{T}_{p,s}$), (c) equivalent shear stress of reaction force due to flow pressure at pillar surface ($\tilde{T}_{p,p}$), and (d) equivalent shear stress of reaction force due to flow shear at bottom surface ($\tilde{T}_{b,s}$), in the space of $\tilde{\delta}_x$ and $\tilde{\delta}_y$ at $Re = 33$.

## IV. Conclusions

The regularly arranged surface structures have some unique effects on the behaviors and transport mechanisms of the fluid flowing above. As an extension of our previous work, this paper focuses on the influence of the arrangement of micro pillars on the basic flow patterns and the complex surface friction.

In this study, the micro pillars are arranged in quadrilateral, and different arrangements are acquired by changing the streamwise and spanwise distances between pillar rows.

The results indicate that the pillar distances in the streamwise and spanwise directions play a key role in the flow evolution and the momentum transport from the upper fluid to the fluid within the spaces among micro pillars. When the streamwise pillar distance is the same as the spanwise distance, the increase in pillar distance from a small value first leads to an increase in eddy size. After a critical point, the further increase in pillar distance causes the decrease in eddy size. During this process, the momentum flux from the upper flow to the flow among micro pillars keeps increasing, and the disturbance to the velocity components around micro pillars becomes stronger. When the streamwise pillar distance is small, the micro eddies are significantly suppressed, and the flow characteristics demonstrate more 2D features, as the functions of spanwise and vertical coordinates. The increase in the spanwise pillar distance enhances the momentum transport from the upper flow to the lower flow. When the spanwise pillar distance is small, the micro eddies connect with each other and form a tubular eddy between each pair of spanwise pillar rows, and the flow characteristics demonstrate more 2D features, as the functions of streamwise and vertical coordinates. The tubular eddies significantly reduce the momentum transport from the upper flow to the lower flow. The increase in the streamwise pillar distance increase the momentum flux slightly.

As another focus of this work, the effect of pillar arrangement on the surface friction was also analyzed. The components of total friction are related to two factors, the dilution effect of the decreasing number density of micro pillars and the multi-faceted effects of micro eddies in the gaps between the streamwise neighboring pillars. These two factors are determined by the streamwise and spanwise pillar distances. Usually, the increase in the streamwise and spanwise pillar distances enhances the momentum transport, so the reaction forces of each single pillar due to flow shear and flow pressure increase, and the equivalent shear stress of the reaction force of bottom plane also increases. However, the dilution effect of decreasing pillar number density decreases the equivalent shear stresses of the reaction forces of micro pillars due to flow shear and flow pressure. The micro eddies also play an important role in determining the reaction forces. The micro eddies reduce the momentum transport from the upper flow to the lower flow and increase the impact of the flow on pillar surfaces. When the spanwise pillar distance is small, the separate micro eddies connect with each other and form a tubular eddy between each pair of spanwise pillar rows. The tubular eddies further reduce the momentum transport and increase the flow impact on pillar surfaces. As a result, the equivalent stress of the pillar reaction force due to flow shear decreases, the equivalent stress of the pillar reaction force due to flow pressure increases, and the equivalent stress of the bottom plane decreases, when the spanwise pillar distance is small and the streamwise pillar distance is comparable to the pillar height.

When the pillar distance is infinitely small, the complex flow will reduce to the flow over a smooth plane at pillar tips, and when the pillar distance becomes infinitely large, the flow will reduce to that over a smooth plane at pillar roots.

**References**

[1] T.C. Hobæk, K.G. Leinan, H.P. Leinaas, C. Thaulow, Surface nanoengineering inspired by evolution, BioNanoSci. **1**, 63 (2011).

[2] D. Patel, V.K. Jain, J. Ramkumar, Micro texturing on metallic surfaces: state of the art, J. Eng. Manuf. **232,** 941 (2018).

[3] J.J. Brandner, E. Anurjew, L. Bohn, E. Hansjosten, T. Henning, U. Schygulia, A. Wenka, K. Schubert, Concepts and realization of microstructure heat exchangers for enhanced heat transfer, Exp. Therm. Fluid Sci. **30**, 801 (2006).

[4] K. Jasch, S. Scholl, Evaluation of heat transfer on micro-structured surfaces based on entropy production, Chem. Eng. Tech. **36**, 993 (2013).

[5] R.H.W. Lam, Y. sun, W. Chen, J. Fu, Elastomeric microposts integrated into microfluidics for flow-mediated endothelial mechanotransduction analysis, Lab Chip. 12, 1865 (2012).

[6] M.T. Frey, I.Y. Tsai, T.P. Russell, S.K. Hanks, Y.L. Wang, cellular responses to substrate topography: role of myosin II and focal adhesion kinase, Biophys. J. 90, 3774 (2006).

[7] K.A. Whitehead, J. Verran, The effect of surface topography on the retention of microorganisms, Food Bioprod. Process. **84**, 253 (2006).

[8] S. Perni, P. Prokopovich, Micropatterning with conical features can control bacterial adhesion on silicone, Soft Matter, **9**, 1844 (2013).

[9] M.V. Graham, A.P. Mosier, T.R. Kiehl, A.E. Kaloyeros, N.C. Cady, Developments of antifouling surfaces to reduce bacterial attachment, Soft Matter **9**, 6235 (2013).

[10] M.A. Burns, B.N. Johnson, S.N. Brahmasandra, K. Handique, J.R. Webster, M. Krishnan, T.S. Sammarco, P.M. Man, D. Jones, D. Heldsinger, C.H. Mastrangelo, D.T. Burke, An integrated nanoliter DNA analysis device, Sci. **282**, 484 (1998).

[11] M. Masaeli, E. Sollier, H. Amini, W. Mao, K. Camacho, N. Doshi, S. Mitragotri, A. Alexeev, D.D. Carlo, Continuous inertial focusing and separation of particles by shape, Phys. Rev. X, **2**, 031017 (2012).

[12] N. Schneider, Exploration of the effect of surface roughness on heat transfer in microscale liquid flow, Ph.D. thesis. Rochester Institute of Technology, 2010.

[13] D.E. Conway, M.T. Breckenridge, E. Hinde, E. Gratton, C.S. Chen, M.A. Schwartz, Fluid shear stress on endothelial cells modulates mechanical tension across VE-cadherin and PECAM-1, Cur. Bio. 23, 1024 (2013).

[14] H.B. Evans, S. Gorumlu, B. Aksak, L. Castillo, J. Sheng, Holographic microscopy and microfluidics platform for measuring wall stress and 3D flow over surfaces textured by micro-pillars, Sci. Rep. **6**, 28753 (2016).

[15] N.S.K. Gunda, J. Joseph, A. Tamayol, M. Akbari, S.K. Mitra, Measurement of pressure drop and flow resistance in microchannels with integrated micropillars, Microfluid Nanofluid, 14, 711 (2013).

[16] Y. Ichikawa, K. Yamamoto, M. Motosuke, Three-dimensional flow velocity and wall shear stress distribution measurement on a micropillar-arrayed surface using astigmatism PTV to understand the influence of microstructures on the flow field, 22, 73 (2018).

[17] Y. Wang, H. Wan, T. Wei, F. Shu, In-depth characterization and analysis of simple shear flows over regularly arranged micro pillars, I. effect of fluid inertia, arXiv:2203.05607.

[18] Y. Wang, J.G. Brasseur, G.G. Banco, A.G. Webb, A.C. Ailiani, T. Neuberger, Development of a lattice-Boltzmann method for multiscale transport and absorption with application to intestinal function, Computational Modeling in Biomechanics, 69, Springer, Dordrecht, 2010.

[19] Y.H. Qian, D. d'Humières, P. Lallemand, Lattice BGK models for Navier-Stokes equation, Europhys. Lett. **17**, 479 (1992).

[20] M. Bouzidi, M. Firdaouss, P. Lallemand, Momentum transfer of a lattice Boltzmann fluid with boundaries, Phys. Fluids **13**, 3452 (2001).

[21] D. Yu, R. Mei, W. Shyy, A multi-block Boltzmann method for viscous fluid flows, Int. J. Num. Meth. Fluids **39**, 99 (2002).

[22] Y. Wang, J.G. Brasseur, G.G. Banco, A.G. Webb, A.C. Ailiani, T. Neuberger, A multiscale lattice Boltzmann model of macro-to micro-scale transport, with applications to gut function, Phil. Trans. R. Soc. A **368**, 2863 (2010).

[23] Y. Wang, J.G. Brasseur, Three-dimensional mechanisms of macro-to-micro-scale transport and absorption enhancement by gut villi motions, Phys. Rev. E **95**, 062412 (2017).

[24] Y. Wang, J.G. Brasseur, Enhancement of mass transfer from particles by local shear-rate and correlations with application to drug dissolution, AIChE J. **65**, e16617 (2019).